\documentclass[prd,aps,nofootinbib,onecolumn,showpacs,10pt]{revtex4}
\usepackage{graphicx,epsfig}

\def\pp{{\prime\prime}}
\def\vp{\varepsilon}

\begin{document}

\title{Covariant Light-Front Approach for $B_c$ transition form factors}
\author{ Wei Wang$^{a}$, Yue-Long Shen$^b$, and Cai-Dian L\"u$^{a,c}$}
\affiliation{
 $^a$ Institute of High Energy Physics, Chinese Academy
of Sciences, Beijing 100049, Peoples'  Republic of China\\
 $^b$Institute of Information Science and Engineering, Ocean University of China, Qingdao 266100, Peoples'  Republic of China\\
 $^c$ Kavli Institute for Theoretical Physics China,  Chinese Academy of Sciences, Beijing 100190,  Peoples'  Republic of China}

\begin{abstract}

In the covariant light-front quark model, we investigate the form
factors of  $B_c$ decays into $D, D^*, D_s, D_s^*, \eta_c, J/\psi,
B, B^*, B_s, B_s^*$ mesons. The form factors in the spacelike region
are directly evaluated. To extrapolate the form factors to the
physical region, we fit the form factors by adopting a suitable
three-parameter form.   At the maximally recoiling point, $b\to
u,d,s$ transition form factors are smaller than $b\to c$ and $c\to
d,s$ form factors, while the $b\to u,d,s,c$ form factors at the zero
recoiling point are close to each other. In the fitting procedure,
we find that parameters in $A_2^{B_cB^*}$ and $A_2^{B_c B^*_s}$
strongly depend on decay constants of $B^*$ and $B_s^*$ mesons.
Fortunately, semileptonic and nonleptonic $B_c$ decays are not
sensitive to these two form factors. We also investigate branching
fractions, polarizations of the semileptonic $B_c$ decays. $B_c\to
(\eta_c,J/\psi)l\nu$ and $B_c\to (B_s,B_s^*)l\nu$ decays have much
larger branching fractions than $B_c\to (D,D^*,B,B^*)l\nu$. For the
three kinds of $B_c\to Vl\nu$ decays, longitudinal  contributions
are  comparable with the transverse contributions. These predictions
will be tested on the ongoing and forthcoming hadron colliders.

\end{abstract}


 \maketitle


\section{Introduction}

$B$ meson decays provide a golden place to extract  magnitudes and
phases  of the Cabibbo-Kobayashi-Maskawa (CKM) matrix elements,
which can test the origins of CP violation in and beyond the
standard model (SM). There has been remarkable progress in the study
of  semileptonic and nonleptonic $B$ meson decays. Experimentally,
the two $B$ factories have accumulated more than $ 10^9$ $B\bar B$
events. Some rare decays with branching fractions of the order
$10^{-7}$ have been observed. On the theoretical
side, great successes have also been achieved: 
apart from contributions proportional to the form factors,  the
so-called nonfactorizable diagrams and some other radiative
corrections are taken into account. All of them make $B$ physics
suitable for the precise test of the SM and the search of new
phenomena (See Ref.~\cite{Buchalla:2008jp} for a recent review).

Compared with $B$ mesons,  $B_c$ meson is heavier: the mass of a
$B_c\bar B_c$ pair has exceeded the threshold of $\Upsilon(4S)$,
thus $B_c$ mesons can not be produced on the $B$ factories. But
$B_c$ meson has a promising prospect on the hadron colliders.  The
Large Hadron Collider (LHC) experiment, which is scheduled to run in
the very near future, will produce plenty of $B_c$ events. With more
data accumulated in the future, the study on $B_c$ mesons will be of
great importance. 
$B_c$ meson can decay not only via the $b\to q$
(q=u,d,s,c) transition like the lighter $B_{u,d,s,}$ mesons, but
also through the $c\to q$ (q=u,d,s) transitions. The CKM matrix
element in the $c\to s$ transition $|V_{cs}|\sim 1$ is much larger
than the CKM matrix element $|V_{cb}|\sim 0.04$ in $b\to c$
transition. Although the phase space in $c\to d,s$ decays is smaller
than that in $b\to c$ transition, the former decays provide about
$70\%$ to the decay width of $B_c$. This results in a larger decay
width and a much smaller lifetime for the $B_c$ meson:
$\tau_{B_c}<\frac{1}{3}\tau_{B}$. The two heavy $b$ and $\bar c$
quarks can annihilate to provide a new kind of weak decays with
sizable partial decay widths. The purely leptonic annihilation decay
$B_c\to l\bar\nu$ can be used to extract the decay constant of $B_c$
and the CKM matrix element $V_{cb}$.

Semileptonic $B_c$ decays are much simpler than nonleptonic decays:
the leptonic part can be straightforwardly evaluated using
perturbation theory leaving only hadronic form factors. In two-body
nonleptonic $B_c$ decays, most channels are also dominated by the
$B_c$ transition form factors. Thus the $B_c$ transition form
factors have already received considerable theoretical
interests~~\cite{Du:1988ws,Colangelo:1992cx,Kiselev:1993ea,Choudhury:1998hs,Kiselev:1999sc,Nobes:2000pm,Ivanov:2000aj,Kiselev:2002vz,Ebert:2003cn,Ivanov:2005fd,
Aliev:2006vs,Hernandez:2006gt,Huang:2007kb,Sun:2008ew,DSV}. In the
present work, we will use the light-front quark model to analyze
these form factors. The light front QCD approach has some unique
features, which are particularly suitable to describe a hadronic
bound state~\cite{Brodsky:1997de}. Based on this approach, a
light-front quark model with many advantages is
developed~\cite{Jaus:1989au,Jaus:1991cy,Cheng:1996if,Choi:2001hg,Jaus:1999zv}.
This model  provides a relativistic treatment of the hadron and also
gives a fully treatment of the hadron spin by using the so-called
Melosh rotation. The light front wave functions, which describe the
hadrons in terms of their fundamental quark and gluon degrees of
freedom, are independent of the hadron momentum and thus are
explicitly Lorentz invariant. In the covariant light-front quark
model~\cite{Jaus:1999zv}, the spurious contribution, which is
dependent on the {orientation} of the light-front, becomes
irrelevant in the study of decay constants and form factors and that
makes the light-front quark model more selfconsistent.  This
covariant model has been successfully extended to investigate the
decay constants and form factors of the  $s$-wave and  $p$-wave
mesons~\cite{Cheng:2003sm,Cheng:2004yj}, the heavy
quarkonium~\cite{Hwang:2006cua}.

Our paper is organized as follows. The formalism of the covariant
light-front quark model is presented in the next section. Numerical
results for the form factors and decay rates of semileptonic $B_c$
decays are given in Section \ref{III}. We also compare our
predictions of form factors with those evaluated in the literature.
Our conclusions are given in Section~\ref{conc}. In the Appendix A,
we give the relation between the form factors defined in various
studies on $B_c$ decays and the widely used Bauer-Stech-Wirbel (BSW)
form factors~\cite{Wirbel:1985ji}. In the Appendix B, we collect
some specific rules when performing the $p^-$ integration.

\section{Covariant light-front quark model}\label{sec:formalism}

$B_c\to P,V$ ($P,V$ denotes a pseudoscalar and a vector meson,
respectively) form factors induced by vector and axial-vector
currents are defined by
  \begin{eqnarray}
     \langle P(P^\pp)| V_\mu | B_c(P')\rangle &=& f_+(q^2) P_\mu +f_-(q^2) q_\mu,\label{eq:formfactor1}\\
     \langle V(P^\pp,\vp^{\pp*})|V_\mu|B_c(P^\prime)\rangle
          &=&\epsilon_{\mu\nu\alpha \beta}\,\vp^{\pp*\nu}P^\alpha q^\beta\,
          g({q^2}),\label{eq:formfactor2}
               \\
    \langle V(P^\pp,\vp^{\pp*})|A_\mu|B_c(P^\prime)\rangle
          &=&-i\left\{\varepsilon_\mu^{\pp*} f({q^2})+\vp^{\pp*}\cdot P
           \left[P_\mu a_+({q^2})
               +q_\mu a_-({q^2})\right]\right\},\label{eq:formfactor3}
\end{eqnarray}
where $P=P^\prime+P^{\prime\prime}$, $q=P^\prime-P^{\prime\prime}$
and the convention $\epsilon_{0123}=1$ is adopted. The vector and
axial-vector currents are defined as $\bar\psi\gamma_\mu\psi'$ and
$\bar\psi\gamma_\mu\gamma_5\psi'$. In $b\to q$ ($q=u,d,s,c$)
transition, $\psi$ and $\psi'$ denotes the $q$ quark field  and the
$b$ quark field, respectively; while in $c\to q'$ ($q'=u,d,s$)
transition, $\psi$ and $\psi'$ denotes the $q'$ quark field  and the
$c$ quark field, respectively. In the literature, the
Bauer-Stech-Wirbel (BSW)~\cite{Wirbel:1985ji} form factors are more
frequently used:
\begin{eqnarray}
       \langle P(P^\pp)|V_\mu|B_c(P^\prime)\rangle &=&
        \left(P_\mu-\frac{m_{B_c}^2-m_P^{2}}{q^2}q_\mu\right) F_1^{B_cP}(q^2)+\frac{m_{B_c}^2-m_P^{2}}{q^2}q_\mu F_0^{B_cP}(q^2) , \\
      \langle V(P^\pp,\vp^{\pp*})|V_\mu|B_c(P^\prime)\rangle &=&
       -\frac{1}{ m_{B_c}+m_{V}}\,\epsilon_{\mu\nu\alpha \beta}\vp^{\pp*\nu}P^\alpha
    q^\beta  V^{B_cV}(q^2),    \\
\ \ \ \
      \langle V(P^\pp,\vp^{\pp*})|A_\mu|
    B_c(P^\prime)\rangle &=& i\Big\{
         (m_{B_c}+m_{V})\vp^{\pp*}_\mu A_1^{B_cV}(q^2)-\frac{\vp^{\pp*}\cdot P}
         { m_{B_c}+m_{V}}\,
         P_\mu A_2^{B_cV}(q^2)    \nonumber \\
    && -2m_{V}\,{\vp^{\pp*}\cdot P\over
    q^2}\,q_\mu\big[A_3^{B_cV}(q^2)-A_0^{B_cV}(q^2)\big]\Big\}.
 \end{eqnarray}
These two kinds of form factors are related to each other via:
 \begin{eqnarray}
 F_1^{B_cP}(q^2)&=&f_+(q^2),\;\;\;
 F_0^{B_cP}(q^2)=f_+(q^2)+\frac{q^2}{m_{B_c}^2-m_P^2}f_-(q^2),\nonumber\\
 V^{B_cV}(q^2)&=&-(m_{B_c}+m_{V})\, g(q^2),\quad
 A_1^{B_cV}(q^2)=-\frac{f(q^2)}{m_{B_c}+m_{V}},\nonumber\\
 A_2^{B_cV}(q^2)&=&(m_{B_c}+m_{V})\, a_+(q^2),\quad
 A_3^{B_cV}(q^2)-A_0^{B_cV}(q^2)=\frac{q^2}{2 m_{V}}\,
 a_-(q^2),\label{eq:relation1}
 \end{eqnarray}
with $A_3^{B_cV}(0)=A_0^{B_cV}(0)$, and
\begin{eqnarray}
 A_3^{B_cV}(q^2)&=&\frac{m_{B_c}+m_{V}}{ 2m_{V}}
 A_1^{B_cV}(q^2)-\frac{m_{B_c}-m_{V}}{2m_{V}}\,A_2^{B_cV}(q^2).\label{eq:relation}
 \end{eqnarray}

\begin{figure}
\includegraphics[scale=0.5]{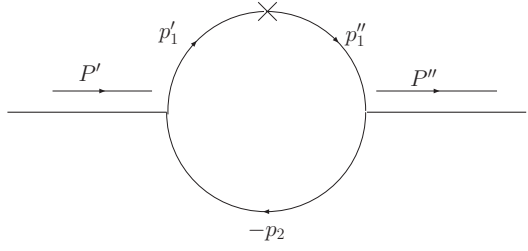}
\caption{Feynman diagram for $B_c\to P,V$ decay amplitudes. The X in
the diagram denotes the vector or axial-vector transition vertex
while the meson-quark-antiquark vertices are given in the
text.}\label{fig:feyn}
\end{figure}

In the covariant light-front quark model, we will work in the
$q^+=0$ frame and employ the light-front decomposition of the
momentum $P^{\prime}=(P^{\prime -}, P^{\prime +}, P^\prime_\bot)$,
where $P^{\prime\pm}=P^{\prime0}\pm P^{\prime3}$, so that $P^{\prime
2}=P^{\prime +}P^{\prime -}-P^{\prime 2}_\bot$. The incoming and
outgoing mesons have the momenta $P^{\prime}=p_1^{\prime}+p_2$ and
$P^{\pp}=p_1^{\pp}+p_2$ and the mass $M^{\prime}$ and
$M^{\prime\prime}$, respectively. For the $B_c$ transition form
factors, $M^{\prime}=m_{B_c}$.  The quark and antiquark inside the
incoming (outgoing) meson have the mass $m_1^{\prime(\pp)}$ and
$m_2$ and the momenta $p_1^{\prime(\pp)}$ and $p_2$, respectively.
These momenta can be expressed in terms of the internal variables
$(x_i, p_\bot^\prime)$ as
 \begin{eqnarray}
 p_{1,2}^{\prime+}=x_{1,2} P^{\prime +},\qquad
 p^\prime_{1,2\bot}=x_{1,2} P^\prime_\bot\pm p^\prime_\bot,
 \end{eqnarray}
with $x_1+x_2=1$. Using these internal variables, one can define
some useful quantities for the incoming meson:
\begin{eqnarray}
 M^{\prime2}_0
          &=&(e^\prime_1+e_2)^2=\frac{p^{\prime2}_\bot+m_1^{\prime2}}
                {x_1}+\frac{p^{\prime2}_{\bot}+m_2^2}{x_2},\quad\quad
                \widetilde M^\prime_0=\sqrt{M_0^{\prime2}-(m^\prime_1-m_2)^2},
 \nonumber\\
 e^{(\prime)}_i
          &=&\sqrt{m^{(\prime)2}_i+p^{\prime2}_\bot+p^{\prime2}_z},\quad\qquad
 p^\prime_z=\frac{x_2 M^\prime_0}{2}-\frac{m_2^2+p^{\prime2}_\bot}{2 x_2
 M^\prime_0},
 \end{eqnarray}
where $e^{(\prime)}_i$ can be interpreted as the energy of the quark
or the antiquark and $M_0^\prime$ can be viewed as the kinematic
invariant mass of the meson system. The definition of the internal
quantities for the outgoing meson is similar. To compute the
hadronic amplitudes, we require the Feynman rules for the
meson-quark-antiquark vertices ($i\Gamma^\prime_M$):
\begin{eqnarray}
i\Gamma_P^\prime&=&
      H^\prime_P\gamma_5,
            \\
i\Gamma_V^\prime&=&iH^\prime_{V}[\gamma_\mu-\frac{1}{W^\prime_{V}}(p^\prime_1-p_2)_\mu].
\end{eqnarray}
For the outgoing meson, one should use $i(\gamma_0
\Gamma^{\prime\dagger}_M\gamma_0)$ for the relevant vertices.

In the conventional light-front quark model, the constituent quarks
are required to be on mass shell and physical quantities can be
extracted from the plus component of the current matrix elements.
However, this framework suffers from the problem of non-covariance
because of the missing zero-mode contributions. In order to solve
this problem, Jaus has proposed the covariant light-front approach
which provides a systematical way to deal with the zero-mode
contributions~\cite{Jaus:1999zv}. Physical quantities such as decay
constants and form factors can be calculated in terms of Feynman
momentum loop integrals which are manifestly covariant. For example,
the lowest order contribution to a form factor is depicted in
Fig.~\ref{fig:feyn} and the $P\to P$ transition amplitude is given
by:
\begin{eqnarray}
 {\cal B}^{PP}_\mu=-i^3\frac{N_c}{(2\pi)^4}\int d^4 p^\prime_1
 \frac{H^\prime_P ( H^\pp_P)}{N_1^\prime N_1^\pp N_2} S^{PP}_{\mu},
\end{eqnarray}
where $N_1^{\prime(\prime\prime)}= p_1^{\prime(\prime\prime)2}
-m_1^{\prime (\prime\prime)2} $, $N_2=p_2^2-m_2^2$.
\begin{eqnarray}
 S^{PP}_\mu &=&{\rm Tr}\left[\gamma_5(\not \!p^\pp_1+m_1^\pp)\gamma_\mu(\not
 \!p^\prime_1+m_1^\prime)\gamma_5(-\not\!p_2+m_2)\right]\nonumber\\
           &=&2 p^\prime_{1\mu} [M^{\prime 2}+M^{\pp2}-q^2-2
           N_2-(m_1^\prime-m_2)^2-(m^\pp_1-m_2)^2+(m_1^\prime-m_1^\pp)^2]
 \nonumber\\
          &&+q_\mu[q^2-2 M^{\prime2}+N^\prime_1-N^\pp_1+2
          N_2+2(m_1^\prime-m_2)^2-(m_1^\prime-m_1^\pp)^2]
 \nonumber\\
          &&+P_\mu[q^2-N^\prime_1-N^\pp_1-(m_1^\prime-m_1^\pp)^2].
 \label{eq:SPPVappen}
\end{eqnarray}

In practice, we use the light-front decomposition of the loop
momentum and perform the integration over the minus component using
the contour method. If the covariant vertex functions are not
singular when performing the integration, the transition amplitude
will pick up the singularities in the anti-quark propagator.  The
integration then leads to:
 \begin{eqnarray}
 N_1^{\prime(\pp)}
      &\to&\hat N_1^{\prime(\pp)}=x_1(M^{\prime(\pp)2}-M_0^{\prime(\pp)2}),
 \nonumber\\
 H^{\prime(\pp)}_M
      &\to& h^{\prime(\pp)}_M,
 \nonumber\\
 W^\pp_M
      &\to& w^\pp_M,
 \nonumber\\
\int \frac{d^4p_1^\prime}{N^\prime_1 N^\pp_1 N_2}H^\prime_P H^\pp_P
S
      &\to& -i \pi \int \frac{d x_2 d^2p^\prime_\bot}
                             {x_2\hat N^\prime_1
                             \hat N^\pp_1} h^\prime_P h^\pp_P \hat S,
 \end{eqnarray}
where
 \begin{eqnarray}
 M^{\pp2}_0
          =\frac{p^{\pp2}_\bot+m_1^{\pp2}}
                {x_1}+\frac{p^{\pp2}_{\bot}+m_2^2}{x_2},
 \end{eqnarray}
with $p^\pp_\bot=p^\prime_\bot-x_2\,q_\bot$. The explicit forms of
$h^\prime_M$ and $w^\prime_M$ for the pseudoscalar and vector meson
are given by:
\begin{eqnarray}
 h^\prime_P&=&h^\prime_V
                  =(M^{\prime2}-M_0^{\prime2})\sqrt{\frac{x_1 x_2}{N_c}}
                    \frac{1}{\sqrt{2}\widetilde M^\prime_0}\varphi^\prime,
 \nonumber\\
                     w^\prime_V&=&M^\prime_0+m^\prime_1+m_2,\label{eqs:wa}
\end{eqnarray}
where $\varphi'$ is the light-front { wave function} for
pseudoscalar and vector mesons. After this integration, the
conventional light-front model is recovered but manifestly the
covariance is lost as it receives additional spurious contributions
proportional to the lightlike four vector $\tilde\omega=(0,2,{\bf
0_{\perp}})$. The undesired spurious contributions can be eliminated
by the inclusion of the zero mode contribution which amounts to
performing the $p^-$ integration in a proper way. The specific rules
under this $p^-$ integration are derived in
Ref.~\cite{Jaus:1999zv,Cheng:2003sm} and the relevant ones in this
work are collected in the Appendix B.

Using Eqs.~(\ref{eq:SPPVappen})--(\ref{eqs:wa}){ and taking the
advantage of the rules in Ref.~\cite{Jaus:1999zv,Cheng:2003sm}}, we
obtain expressions for the $P\to P$ form factors:
\begin{eqnarray}
 f_+(q^2)&=&\frac{N_c}{16\pi^3}\int dx_2 d^2p^\prime_\bot
            \frac{h^\prime_P h^\pp_P}{x_2 \hat N_1^\prime \hat N^\pp_1}
            \bigg[x_1 (M_0^{\prime2}+M_0^{\pp2})+x_2 q^2
 \nonumber\\
         &&\qquad-x_2(m_1^\prime-m_1^\pp)^2 -x_1(m_1^\prime-m_2)^2-x_1(m_1^\pp-m_2)^2\bigg],
 \nonumber\\
 f_-(q^2)&=&\frac{N_c}{16\pi^3}\int dx_2 d^2p^\prime_\bot
            \frac{2h^\prime_P h^\pp_P}{x_2 \hat N_1^\prime \hat N^\pp_1}
            \Bigg\{- x_1 x_2 M^{\prime2}-p_\bot^{\prime2}-m_1^\prime m_2
                  +(m_1^\pp-m_2)(x_2 m_1^\prime+x_1 m_2)
\nonumber\\
         &&\qquad +2\frac{q\cdot P}{q^2}\left(p^{\prime2}_\bot+2\frac{(p^\prime_\bot\cdot q_\bot)^2}{q^2}\right)
                  +2\frac{(p^\prime_\bot\cdot q_\bot)^2}{q^2}
                  -\frac{p^\prime_\bot\cdot q_\bot}{q^2}
                  \Big[M^{\pp2}-x_2(q^2+q\cdot P)
\nonumber\\
         &&\qquad -(x_2-x_1) M^{\prime2}+2 x_1 M_0^{\prime
                  2}-2(m_1^\prime-m_2)(m_1^\prime+m_1^\pp)\Big]
           \Bigg\}.
 \label{eq:fpm}
\end{eqnarray}
Similarly, the $P\to V$ transition amplitudes are given by:
 \begin{eqnarray}
 {\cal B}^{PV}_\mu=-i^3\frac{N_c}{(2\pi)^4}\int d^4 p^\prime_1
 \frac{H^\prime_P (i H^\pp_V)}{N_1^\prime N_1^\pp N_2} S^{PV}_{\mu\nu}\,\vp^{\pp*\nu},
 \end{eqnarray}
where
 \begin{eqnarray}
S^{PV}_{\mu\nu} &=&(S^{PV}_{V}-S^{PV}_A)_{\mu\nu}
 \nonumber\\
                &=&{\rm Tr}\left[\left(\gamma_\nu-\frac{1}{W^\pp_V}
                (p_1^\pp-p_2)_\nu\right)
                                 (\not \!p^\pp_1+m_1^\pp)
                                 (\gamma_\mu-\gamma_\mu\gamma_5)
                                 (\not \!p^\prime_1+m_1^\prime)\gamma_5(-\not
                                 \!p_2+m_2)\right]\nonumber \\
           &=&-2 i\epsilon_{\mu\nu\alpha\beta}
                 \Big\{p^{\prime\alpha}_1 P^\beta (m_1^\pp-m_1^\prime)
                   +p^{\prime\alpha}_1q^\beta(m_1^\pp+m_1^\prime-2
                   m_2)+q^\alpha P^\beta m_1^\prime
                 \Big\}
 \nonumber\\
           &&+\frac{1}{W^\pp_V}(4 p^\prime_{1\nu}-3
           q_\nu-P_\nu)i\epsilon_{\mu\alpha\beta\rho}p^{\prime\alpha}_1 q^\beta P^\rho
 \nonumber\\
           &&+2
           g_{\mu\nu}\Big\{m_2(q^2-N_1^\prime-N^\pp_1-m_1^{\prime2}-m_1^{\pp2})
                      -m_1^\prime (M^{\pp2}-N_1^\pp-N_2-m_1^{\pp2}-m_2^2)
 \nonumber\\
           &&-m^\pp_1(M^{\prime2}-N^\prime_1-N_2-m_1^{\prime2}-m_2^2)
           -2 m_1^\prime m_1^\pp m_2\Big\}
  \nonumber\\
           &&+8 p^\prime_{1\mu} p^\prime_{1\nu}(m_2-m_1^\prime)
             -2(P_\mu q_\nu+q_\mu P_\nu+2q_\mu q_\nu) m_1^\prime
             +2p^\prime_{1\mu} P_\nu (m_1^\prime-m_1^\pp)
 \nonumber\\
           &&+2 p^\prime_{1\mu} q_\nu(3 m_1^\prime-m_1^\pp-2m_2)+2
           P_\mu p^\prime_{1\nu}(m_1^\prime+m_1^\pp)+2 q_\mu
           p^\prime_{1\nu}(3 m_1^\prime+m_1^\pp-2 m_2)
 \nonumber\\
           &&+\frac{1}{2W^\pp_V}(4 p^\prime_{1\nu}-3q_\nu-P_\nu)
              \Big\{2 p^\prime_{1\mu}[M^{\prime2}+M^{\pp2}-q^2
                -2 N_2+2(m_1^\prime-m_2)(m_1^\pp+m_2)]
 \nonumber\\
           &&   +q_\mu[q^2-2 M^{\prime2}+N^\prime_1-N_1^\pp+2 N_2-(m_1+m_1^\pp)^2
           +2(m_1^\prime-m_2)^2]
 \nonumber\\
           &&   +P_\mu[q^2-N_1^\prime-N_1^\pp-(m_1^\prime+m_1^\pp)^2]
              \Big\}.
              \label{eq:BtoV}
 \end{eqnarray}
The above equations give the  expression for $P\to V$ form factors:
\begin{eqnarray}
 g(q^2)&=&-\frac{N_c}{16\pi^3}\int dx_2 d^2 p^\prime_\bot
           \frac{2 h^\prime_P h^\pp_V}{x_2 \hat N^\prime_1 \hat N^\pp_1}
           \Bigg\{x_2 m_1^\prime+x_1 m_2+(m_1^\prime-m_1^\pp)
           \frac{p^\prime_\bot\cdot q_\bot}{q^2}
           +\frac{2}{w^\pp_V}\left[p^{\prime2}_\bot+\frac{(p^\prime_\bot\cdot
            q_\bot)^2}{q^2}\right]
           \Bigg\},
  \nonumber\\
  f(q^2)&=&\frac{N_c}{16\pi^3}\int dx_2 d^2 p^\prime_\bot
            \frac{ h^\prime_P h^\pp_V}{x_2 \hat N^\prime_1 \hat N^\pp_1}
            \Bigg\{2
            x_1(m_2-m_1^\prime)(M^{\prime2}_0+M^{\pp2}_0)-4 x_1
            m_1^\pp M^{\prime2}_0+2x_2 m_1^\prime q\cdot P
  \nonumber\\
         &&+2 m_2 q^2-2 x_1 m_2
           (M^{\prime2}+M^{\pp2})+2(m_1^\prime-m_2)(m_1^\prime+m_1^\pp)^2
           +8(m_1^\prime-m_2)\left[p^{\prime2}_\bot+\frac{(p^\prime_\bot\cdot
            q_\bot)^2}{q^2}\right]
  \nonumber\\
         &&
           +2(m_1^\prime+m_1^\pp)(q^2+q\cdot
           P)\frac{p^\prime_\bot\cdot q_\bot}{q^2}
           -4\frac{q^2 p^{\prime2}_\bot+(p^\prime_\bot\cdot q_\bot)^2}{q^2 w^\pp_V}
            \bigg[2 x_1 (M^{\prime2}+M^{\prime2}_0)-q^2-q\cdot P
 \nonumber\\
         &&-2(q^2+q\cdot P)\frac{p^\prime_\bot\cdot
            q_\bot}{q^2}-2(m_1^\prime-m_1^\pp)(m_1^\prime-m_2)
            \bigg]\Bigg\},
  \nonumber\\
 a_+(q^2)&=&\frac{N_c}{16\pi^3}\int dx_2 d^2 p^\prime_\bot
            \frac{2 h^\prime_P h^\pp_V}{x_2 \hat N^\prime_1 \hat N^\pp_1}
            \Bigg\{(x_1-x_2)(x_2 m_1^\prime+x_1 m_2)-[2x_1
            m_2+m_1^\pp+(x_2-x_1)
            m_1^\prime]\frac{p^\prime_\bot\cdot q_\bot}{q^2}
  \nonumber\\
         &&-2\frac{x_2 q^2+p_\bot^\prime\cdot q_\bot}{x_2 q^2
            w^\pp_V}\Big[p^\prime_\bot\cdot p^\pp_\bot+(x_1 m_2+x_2 m_1^\prime)(x_1 m_2-x_2
            m_1^\pp)\Big]\Bigg\},
  \nonumber\\
 a_-(q^2)&=&\frac{N_c}{16\pi^3}\int dx_2 d^2 p^\prime_\bot
            \frac{ h^\prime_P h^\pp_V}{x_2 \hat N^\prime_1 \hat N^\pp_1}
            \Bigg\{2(2x_1-3)(x_2 m_1^\prime+x_1 m_2)-8(m_1^\prime-m_2)
            \left[\frac{p^{\prime2}_\bot}{q^2}+2\frac{(p^\prime_\bot\cdot q_\bot)^2}{q^4}\right]
 \nonumber\\
         &&-[(14-12 x_1) m_1^\prime-2 m_1^\pp-(8-12 x_1) m_2]\frac{p^\prime_\bot\cdot q_\bot}{q^2}\nonumber\\
         &&+\frac{4}{w^\pp_V}\bigg([M^{\prime2}+M^{\pp2}-q^2+2(m_1^\prime-m_2)(m_1^\pp+m_2)]
                                   (A^{(2)}_3+A^{(2)}_4-A^{(1)}_2)
 \nonumber\\
         &&+Z_2(3 A^{(1)}_2-2A^{(2)}_4-1)+\frac{1}{2}[x_1(q^2+q\cdot P)
            -2 M^{\prime2}-2 p^\prime_\bot\cdot q_\bot -2 m_1^\prime(m_1^\pp+m_2)
 \nonumber\\
         &&-2 m_2(m_1^\prime-m_2)](A^{(1)}_1+A^{(1)}_2-1)
         q\cdot P\Bigg[\frac{p^{\prime2}_\bot}{q^2} +\frac{(p^\prime_\bot\cdot q_\bot)^2}{q^4}\Bigg] (4A^{(1)}_2-3)\bigg)
            \Bigg\}.
 \end{eqnarray}
The functions $Z_2$ and $A^{(1)}_1$, $A^{(1)}_2$, $A^{(2)}_3$,
$A^{(2)}_4$, and~$Z_2$ are given in the Appendix B. Expressions for
the BSW form factors can be directly obtained through the simple
relation given in Eq.~(\ref{eq:relation1}).

\section{Numerical results}
\label{III}

The $\bar qq$ meson state is described by the light-front wave
function which can be obtained by solving the relativistic
Schr\"odinger equation. But in fact except for some limited cases,
the exact solution is not obtainable. In practice, we usually prefer
to employ a phenomenological wave function to describe the hadronic
structure. In this work, we will use the simple Gaussian-type wave
function which has been extensively examined in the literature:
\begin{eqnarray} \label{eq:Gauss}
 \varphi^\prime
    &=&\varphi^\prime(x_2,p^\prime_\perp)
             =4 \left({\pi\over{\beta^{\prime2}}}\right)^{3\over{4}}
               \sqrt{{dp^\prime_z\over{dx_2}}}~{\rm exp}
               \left(-{p^{\prime2}_z+p^{\prime2}_\bot\over{2 \beta^{\prime2}}}\right),
\nonumber\\
%
         \frac{dp^\prime_z}{dx_2}&=&\frac{e^\prime_1 e_2}{x_1 x_2 M^\prime_0}.
 \label{eq:wavefn}
\end{eqnarray}
The parameter $\beta'$, which describes the momentum distribution,
is expected to be of order $\Lambda_{\rm QCD}$. It is usually fixed
by meson's decay constant whose analytic expression in the covariant
light-front model is given in \cite{Cheng:2003sm}. The decay
constant of $f_{J/\psi}$ can be determined by the leptonic decay
width
\begin{eqnarray}
\Gamma_{ee}\equiv\Gamma(J/\psi\to e^+e^-)=\frac{4\pi
\alpha_{em}^2Q_c^2 f_{J/\psi}^2}{3m_{J/\psi}},
\end{eqnarray}
where $Q_c=+2/3$ denotes the electric charge of the charm quark.
Using the measured results for the electronic width of
$J/\psi$~\cite{Amsler:2008zz}:
\begin{eqnarray}
 \Gamma_{ee}=(5.55\pm0.14\pm0.02) {\rm keV},
\end{eqnarray}
we obtain $f_{J/\psi}=(416\pm5)$~MeV.  Under the factorization
assumption, the decay constant of $\eta_c$ has been extracted by
CLEO collaboration from $B\to\eta_c K$ decays \cite{Edwards:2000bb}:
\begin{eqnarray}
 f_{\eta_c}=(335\pm75) {\rm MeV},
\end{eqnarray}
where the central value is about $20\%$ smaller than that of
$J/\psi$. In this work, we will assume the same decay constant for
$\eta_c$ as that of $J/\psi$. We also introduce an uncertainty of
$20\%$ to this value. Decay constants for charged pseudoscalars are
usually derived through the purely leptonic decays:
\begin{eqnarray}
 \Gamma(P\to l\bar\nu)=\frac{G_F^2 |V_{\rm CKM}|^2}{8\pi }f_P^2m_l^2
 m_P(1-\frac{m_l^2}{m_P^2})^2.
\end{eqnarray}
The experimental results for the decay constants of charmed mesons
are averaged as~\cite{Rosner:2008yu}:
\begin{eqnarray}
 f_{D_s}=(273\pm10) {\rm MeV},\;\;\; f_{D}=(205.8\pm8.9) {\rm MeV}.
\end{eqnarray}
As clearly shown in the above equation, the uncertainties for these
decay constants are less than $5\%$. It provides a solid foundation
for the precise study on $B_c$ transition form factors. In the heavy
quark limit, the decay constant $f_{D^*}$ of a vector heavy meson
$D^*$ is related to that of a pseudoscalar meson through:
\begin{eqnarray}
 f_{D^*}=f_D\times \sqrt{\frac{m_D}{m_{D^*}}},
\end{eqnarray}
where $m_D$ and $m_{D^*}$ denotes the mass of the pseudoscalar and
vector meson, respectively. That implies $f_{D^*}<f_D$ since
$m_{D^*}>m_D$. In the following, we will use the same values for the
decay constant of the vectors and pseudoscalars. To compensate the
differences, we will also introduce an uncertainty of $10\%$ to the
decay constants. Decay constants for the bottom mesons are employed
by:
\begin{eqnarray}
 f_{B_c}=(400\pm40) {\rm MeV},\;\;\; f_B=(190\pm20)  {\rm MeV},\;\;\; f_{B_s}=(230\pm20) {\rm MeV}.
\end{eqnarray}
These values  are slightly smaller than results provided by Lattice
QCD \cite{Chiu:2007km}:
\begin{eqnarray}
 f_{B_s}=(253\pm8\pm7){\rm MeV},\;\;\; f_{B_c}=(489\pm4\pm3) {\rm
 MeV}.
\end{eqnarray}
Decay constants of the vector $B$ mesons are used as:
$f_{B^*}=(210\pm 20)$ MeV and $f_{B_s^*}=(260\pm 20)$ MeV which are
about $10\%$ larger than those of the pseudoscalar $B$ mesons. Shape
parameters $\beta'$s determined from these decay constants, together
with the constituent quark masses used in the calculation, are shown
in table~\ref{tab:input}. The consistent quark masses are close to
the ones used in Ref.~\cite{Cheng:2003sm,Cheng:2004yj}. To estimate
the uncertainties caused by these quark masses, we will introduce
the uncertainties of $0.03$ GeV and $0.1$ GeV  to the light quark
masses and the heavy quark masses, respectively.  The masses (in
units of GeV) of hadrons are used as~\cite{Amsler:2008zz}:
\begin{eqnarray}
 &&m_{B_c}=6.276,\;\;\; m_{D}=1.8645,\;\;\;
 m_{D^*}=2.0067,\;\;\;m_{D_s}=1.9682,\;\;\;m_{D_s^*}=2.112,\nonumber\\
 && m_{\eta_c}=2.9804,\;\;\;m_{J/\psi}=3.0969,\;\;\;m_{B}=5.279,\;\;\;
 m_{B^*}=5.325,\;\;\;m_{B_s}=5.3675,\nonumber\\
 && m_{B_s^*}=m_{B^*}+m_{B_s}-m_{B}.
\end{eqnarray}

\begin{table}
\caption{Input parameters $m_q$ and $\beta'$ (in units of GeV) in
the Gaussian-type light-front wave function (\ref{eq:wavefn}).
Uncertainties of $\beta'$ are from the decay constants as discussed
in the text.}\label{tab:input}
\begin{tabular}{cccccc}
  \hline\hline
   $m_{u,d}$        & $ m_s$        & $m_c$         & $m_b$             \\ \hline
   $0.25$           & $0.37$        & $1.4$         & $4.8$             \\ \hline
   $\beta'_{D}$      & $\beta'_{D^*}$ & $\beta'_{D_s}$ & $\beta'_{D_s^*}$   \\ \hline
   $0.466_{-0.021}^{+0.022}$          & $0.366_{-0.010}^{+0.010}$       & $0.600_{-0.025}^{+0.026}$       & $0.438_{-0.010}^{+0.010}$           \\ \hline
   $\beta'_{B}$      & $\beta'_{B^*}$ & $\beta'_{B_s}$ & $\beta'_{B_s^*}$   \\ \hline
   $0.555_{-0.048}^{+0.048}$          & $0.528_{-0.034}^{+0.033}$       & $0.626_{-0.045}^{+0.045}$       & $0.599_{-0.032}^{+0.033}$           \\ \hline
   $\beta'_{\eta_c}$ & $\beta'_{J/\psi}$              & $\beta'_{B_c}$     \\\hline
   $0.814_{-0.086}^{+0.092}$          & $0.632_{-0.005}^{+0.005}$                       & $0.890^{+0.075}_{-0.074}$           \\\hline\hline
\end{tabular}
\end{table}

If a light meson is emitted in exclusive nonleptonic decays, only
the form factor at maximally recoiling point ($q^2\simeq 0$) is
required but the $q^2$-dependent behavior in the full $q^2>0$ region
is required in semileptonic $B_c$ decays. Because of the condition
$q^+=0$ imposed during the course of calculation, form factors can
be directly studied only at spacelike momentum transfer
$q^2=-q^2_\bot\leq 0$, which are not relevant for the semileptonic
processes. It has been proposed in \cite{Cheng:2003sm} to
parameterize form factors as explicit functions of $q^2$ in the
space-like region and one can analytically extend them to the
time-like region. To shed light on the momentum dependence, we will
choose the parametrization for the $b$ quark decays:
\begin{eqnarray}
 F(q^2)=F(0){\rm exp}(c_1\hat s+c_2\hat s^2),
\end{eqnarray}
where $\hat s=q^2/m_{B_c}^2$ and $F$ denotes anyone of the form
factors $F_1,F_0$ and $V,A_0,A_1,A_2$.  But for $c\to u,d,s$
transitions, we find that the fitted values for the two parameters
$c_1,c_2$ are not stable and thus we adopt the optional
three-parameter form:
\begin{eqnarray}
 F(q^2)=\frac{F(0)}{1-\frac{q^2}{m_{fit}^2}+\delta(\frac{q^2}{m_{fit}^2})^2}.
\end{eqnarray}

In the procedure to fit the form factors $A_2^{B_c B^*}$ and
$A_2^{B_cB_s^*}$, we find that the shape parameters
($m_{fit},\delta$) strongly depend on the decay constants $f_{B^*}$
and $f_{B_s^*}$. In this case, our predictions on these two form
factors are  unreliable, thus we refrain from predicting these two
form factors. Fortunately, the ambiguity of $A_2^{B_c B^*}$ and
$A_2^{B_cB_s^*}$ will not affect the physical quantities in various
physical decay channels. As we can see from
equation~(\ref{eq:relation}), the masses of $B^*,B_s^*$ mesons are
very close to that of $B_c$, thus the second term in the right hand
side is negligible. The form factor $A_0$, which is relevant for the
nonleptonic $B_c\to B^*(B_s^*)P$ decays, receives small
contributions from $A_2$. Contributions from $A_2$ to the $B_c\to
B^*(B_s^*)l\nu$ decays and $B_c\to B^*(B_s^*)V$ decays are also
small which will be shown in the following.

Our predictions of the remanent form factors are collected in
table~\ref{tab:resultsformfactors1} and
table~\ref{tab:resultsformfactors2}. The first kind of uncertainties
shown in these tables are from those in decay constants of the $B_c$
meson and the final mesons; while the second kind of uncertainties
are from those in the constituent quark masses. Several remarks are
given in order. First, from these two tables, we can see that the
$B_c\to D,D^*,D_s,D_s^*$ form factors at maximally recoiling point
($q^2=0$) are smaller than the other ones. It can be understood as
follows. In $B_c\to D,D^*,D_s,D_s^*$ transitions, the initial charm
quark is almost at rest and its momentum is of order $m_c$; in the
final state, the meson moves very fast and the charm quark tends to
have a very large momentum of order $m_b$. In this transition, the
overlap between the wave functions is limited which will produce
small values for the form factors. In $B_c\to \eta_c, J/\psi$
transitions, the spectator charm antiquark in $\eta_c,J/\psi$ play
the same role with the charm quark generated from the weak vertex.
The light-front wave function of the charmonium is expected to have
a maximum at $E=\frac{m_{B_c}^2+m_{\eta_c}^2}{4m_{B_c}}\sim
\frac{m_{B_c}}{4}\approx m_c$. The overlap between the initial and
final states' light-front wave functions in $B_c\to\eta_c, J/\psi$
becomes larger, which certainly induces larger form factors. It is
also similar for the $B_c\to B,B_s$ form factors. Secondly, the
$B_c\to D_s,\eta_c$ form factors at the zero recoiling point are
close to each other. The initial charm quark is almost at rest and
its momentum is of order $m_c$. In these two kinds of transitions,
the charm spectator in the final states tends to posses a momentum
of order $m_c$. The overlaps of the wave functions in $B_c\to
D_s,\eta_c$ transitions are expected to be in similar size. Thirdly,
the SU(3) symmetry breaking effects in $B_c\to D,D_s$ and $B_c\to
D^*,D_s^*$ form factors are quite large, as the decay constant of
$D_s$ is about one third larger than that of the $D$ meson. But in
$B_c\to B,B_s$ and $B_c\to B^*,B_s^*$ transitions, the SU(3)
breaking effect is small, because the decay constants
$f_{B^{(*)},B_s^{(*)}}$ are in similar size. Fourthly,  since the
uncertainties from decay constants of $D,D_s, J/\psi$ are very
small, the relevant uncertainties to the form factors are also very
small.

\begin{table}
\caption{$B_c\to D,D^*,D_s,D_s^*,\eta_c,J/\psi$ form factors in the
light-front quark model. The uncertainties are from the $B_c$ decay
constants and the decay constant of the final state mesons.
}\label{tab:resultsformfactors1}
\begin{tabular}{ccccccccccc}
 \hline\hline
 & $F$ & $F(0)$ &$F(q^2_{\rm {max}})$ & $c_1$ & $c_2$ \\
 \hline
 & $F_1^{B_c D}$ & $0.16^{+0.02+0.02}_{-0.02-0.01}$& $1.10^{+0.07+0.11}_{-0.07-0.10}$ & $3.46^{+0.24+0.19}_{-0.22-0.19}$ &$0.90^{+0.05+0.06}_{-0.05-0.06}$
 \\
 & $F_0^{B_c D}$ & $0.16^{+0.02+0.02}_{-0.02-0.01}$  & $0.59^{+0.02+0.05}_{-0.02-0.05}$   & $2.41^{+0.22+0.17}_{-0.20-0.17}$  & $0.47^{+0.04+0.04}_{-0.04-0.04}$
 \\
 & $V^{B_c D^*}$ & $0.13^{+0.01+0.02}_{-0.02-0.02}$  & $1.16^{+0.08+0.16}_{-0.07-0.14}$   &  $4.21^{+0.30+0.25}_{-0.27-0.25}$ & $1.09^{+0.07+0.07}_{-0.06-0.07}$
 \\
 & $A_0^{B_c D^*}$ & $0.09^{+0.01+0.01}_{-0.01-0.01}$  & $0.79^{+0.06+0.09}_{-0.05-0.08}$   & $4.18^{+0.30+0.27}_{-0.27-0.27}$  & $0.96^{+0.06+0.08}_{-0.07-0.06}$
 \\
 & $A_1^{B_c D^*}$ & $0.08^{+0.01+0.01}_{-0.01-0.01}$  & $0.42^{+0.02+0.05}_{-0.01-0.04}$   & $3.18^{+0.28+0.24}_{-0.25-0.23}$  & $0.65^{+0.06+0.06}_{-0.04-0.05}$
 \\
 & $A_2^{B_c D^*}$ & $0.07^{+0.01+0.01}_{-0.01-0.01}$  & $0.51^{+0.01+0.07}_{-0.01-0.06}$   & $3.78^{+0.26+0.23}_{-0.24-0.23}$  & $0.80^{+0.04+0.06}_{-0.04-0.05}$\\
 \hline\hline
 & $F_1^{B_c D_s}$ & $0.28^{+0.02+0.02}_{-0.02-0.02}$  & $1.24^{+0.04+0.09}_{-0.05-0.09}$   & $2.78^{+0.17+0.14}_{-0.16-0.14}$  & $0.72^{+0.03+0.04}_{-0.03-0.04}$
 \\
 & $F_0^{B_c D_s}$ & $0.28^{+0.02+0.02}_{-0.02-0.02}$  & $0.68^{+0.01+0.04}_{-0.01-0.04}$   & $1.72^{+0.15+0.12}_{-0.14-0.12}$  & $0.27^{+0.05+0.02}_{-0.05-0.02}$\\
 & $V^{B_c D^*_s}$ & $0.23^{+0.02+0.03}_{-0.02-0.02}$  & $1.36^{+0.07+0.16}_{-0.07-0.14}$   &  $3.63^{+0.23+0.21}_{-0.21-0.21}$ & $0.95^{+0.04+0.06}_{-0.04-0.06}$
 \\
 & $A_0^{B_c D^*_s}$ & $0.17^{+0.01+0.01}_{-0.01-0.01}$  & $0.94^{+0.06+0.08}_{-0.05-0.08}$   & $3.58^{+0.23+0.22}_{-0.21-0.23}$  & $0.83^{+0.06+0.06}_{-0.04-0.06}$\\
 & $A_1^{B_c D^*_s}$ & $0.14^{+0.01+0.02}_{-0.01-0.01}$  & $0.51^{+0.01+0.04}_{-0.01-0.04}$   & $2.62^{+0.21+0.19}_{-0.19-0.19}$  & $0.53^{+0.03+0.04}_{-0.03-0.04}$
 \\
 & $A_2^{B_c D^*_s}$ & $0.12^{+0.01+0.02}_{-0.01-0.02}$  & $0.57^{+0.01+0.06}_{-0.02-0.06}$   & $3.18^{+0.19+0.18}_{-0.18-0.18}$  & $0.66^{+0.03+0.04}_{-0.04-0.04}$\\
 \hline\hline
 & $F_1^{B_c \eta_c}$ & $0.61^{+0.03+0.01}_{-0.04-0.01}$  & $1.09^{+0.00+0.05}_{-0.02-0.05}$   & $1.99^{+0.22+0.08}_{-0.20-0.08}$  & $0.44^{+0.05+0.02}_{-0.05-0.02}$
 \\
 & $F_0^{B_c \eta_c}$ & $0.61^{+0.03+0.01}_{-0.04-0.01}$  & $0.86^{+0.02+0.04}_{-0.03-0.04}$   & $1.18^{+0.26+0.09}_{-0.24-0.09}$  & $0.17^{+0.09+0.02}_{-0.09-0.02}$\\
 & $V^{B_c J/\psi}$ & $0.74^{+0.01+0.03}_{-0.01-0.03}$  & $1.45^{+0.03+0.09}_{-0.04-0.08}$   &  $2.46^{+0.13+0.10}_{-0.13-0.10}$ & $0.56^{+0.02+0.03}_{-0.03-0.03}$
 \\
 & $A_0^{B_c J/\psi}$ & $0.53^{+0.01+0.02}_{-0.01-0.02}$  & $1.02^{+0.02+0.07}_{-0.02-0.07}$   & $2.39^{+0.13+0.11}_{-0.13-0.11}$  & $0.50^{+0.02+0.02}_{-0.03-0.02}$\\
 & $A_1^{B_c J/\psi}$ & $0.50^{+0.01+0.02}_{-0.02-0.02}$  & $0.80^{+0.00+0.05}_{-0.01-0.05}$   & $1.73^{+0.12+0.12}_{-0.12-0.12}$  & $0.33^{+0.01+0.02}_{-0.02-0.02}$
 \\
 &$A_2^{B_c J/\psi}$ & $0.44^{+0.02+0.02}_{-0.03-0.02}$  & $0.81^{+0.02+0.05}_{-0.03-0.04}$   & $2.22^{+0.11+0.11}_{-0.10-0.11}$  & $0.45^{+0.01+0.02}_{-0.01-0.02}$\\
 \hline\hline
\end{tabular}
\end{table}

\begin{table}
\caption{Results for the $B_c\to B,B^*,B_s,B_s^*$ form factors in
the light front quark model.  The uncertainties are from the $B_c$
decay constants and the decay constant of the final state
mesons.}\label{tab:resultsformfactors2}
\begin{tabular}{ccccccccccc}
 \hline\hline
 & $F$ & $F(0)$ &$F(q^2_{\rm {max}})$ & $m_{fit}$ & $\delta$  \\
 \hline
 & $F_1^{B_c B}$ & $0.63^{+0.04+0.03}_{-0.05-0.03}$  & $0.96^{+0.05+0.08}_{-0.07-0.07}$   & $1.19^{+0.09+0.01}_{-0.09-0.01}$  & $0.33^{+0.04+0.01}_{-0.04-0.01}$
 \\
 & $F_0^{B_c B}$ & $0.63^{+0.04+0.03}_{-0.05-0.03}$  & $0.81^{+0.02+0.06}_{-0.03-0.05}$   & $1.52^{+0.22+0.02}_{-0.19-0.02}$  & $0.52^{+0.16+0.02}_{-0.10-0.02}$\\
 & $V^{B_c B^*}$ & $3.29^{+0.17+0.32}_{-0.21-0.30}$  & $4.89^{+0.19+0.61}_{-0.27-0.53}$   &  $2.65^{+0.13+0.05}_{-0.14-0.06}$ & $1.75^{+0.27+0.10}_{-0.22-0.11}$
 \\
 & $A_0^{B_c B^*}$ & $0.47^{+0.01+0.04}_{-0.01-0.04}$  & $0.68^{+0.01+0.07}_{-0.02-0.07}$   & $0.99^{+0.04+0.04}_{-0.04-0.04}$  & $0.31^{+0.03+0.02}_{-0.03-0.02}$\\
 & $A_1^{B_c B^*}$ & $0.43^{+0.01+0.04}_{-0.01-0.04}$  & $0.57^{+0.00+0.06}_{-0.01-0.06}$   & $1.16^{+0.07+0.03}_{-0.07-0.03}$  & $0.27^{+0.03+0.01}_{-0.03-0.02}$  \\
 \hline\hline
 & $F_1^{B_c B_s}$ & $0.73^{+0.03+0.03}_{-0.04-0.03}$  & $1.01^{+0.02+0.07}_{-0.04-0.06}$   & $1.35^{+0.07+0.01}_{-0.08-0.01}$  & $0.35^{+0.04+0.00}_{-0.04-0.01}$
 \\
 & $F_0^{B_c B_s}$ & $0.73^{+0.03+0.03}_{-0.04-0.03}$  & $0.87^{+0.00+0.05}_{-0.02-0.05}$   & $1.77^{+0.24+0.04}_{-0.20-0.04}$  & $0.60^{+0.23+0.04}_{-0.14-0.04}$\\
 & $V^{B_c B_s^*}$ & $3.62^{+0.12+0.31}_{-0.15-0.29}$  & $4.93^{+0.14+0.53}_{-0.19-0.47}$   &  $2.94^{+0.11+0.04}_{-0.11-0.05}$ & $1.78^{+0.25+0.07}_{-0.21-0.08}$
 \\
 & $A_0^{B_c B_s^*}$ & $0.56^{+0.00+0.04}_{-0.01-0.04}$  & $0.75^{+0.00+0.07}_{-0.01-0.07}$   & $1.13^{+0.03+0.04}_{-0.04-0.04}$  & $0.33^{+0.03+0.02}_{-0.03-0.02}$\\
 & $A_1^{B_c B_s^*}$ & $0.52^{+0.00+0.04}_{-0.01-0.04}$  & $0.64^{+0.00+0.06}_{-0.01-0.06}$   & $1.33^{+0.07+0.03}_{-0.07-0.03}$  & $0.28^{+0.03+0.01}_{-0.03-0.01}$  \\
 \hline\hline
\end{tabular}
\end{table}

In the literature, there already exist lots of studies on $B_c$
transition form
factors~\cite{Du:1988ws,Colangelo:1992cx,Kiselev:1993ea,Choudhury:1998hs,Kiselev:1999sc,Nobes:2000pm,Ivanov:2000aj,Kiselev:2002vz,Ebert:2003cn,Ivanov:2005fd,
Aliev:2006vs,Hernandez:2006gt,Huang:2007kb,Sun:2008ew,DSV} and their
results are collected in table \ref{tab:formfactorsD} and table
\ref{tab:formfactorcharmonium}. Since $J/\psi$ can be easily
reconstructed by a lepton pair on the hadron collider, the $B_c\to
J/\psi$ form factors have been widely studied in many theoretical
frameworks. In a very recent paper~\cite{Sun:2008ew}, the authors
have derived two kinds of wave functions for the charmonium state
under harmonic oscillator potential and Coulomb potential. They also
used these wave functions to investigate the $B_c\to \eta_c, J/\psi$
form factors under the perturbative QCD approach. Compared with
their results, our predictions are typically smaller. The main
reason is that they have used a much larger decay constant
$f_{B_c}$. Regardless of this effect, our results are consistent
with theirs.  Results collected in \ref{tab:formfactorsD} (including
ours) have large differences which can be discriminated by the
future LHC experiments. The $B_c\to D_s,D_s^*$ is described as the
FCNC $b\to s$ transition at the quark level which is purely loop
effects in the SM. As a consequence, this transition has a very
small Wilson coefficient and the $B_c\to D_s,D_s^*$ form factors are
less studied in the literature. Similar with the $b\to u,s,c$
transitions, predictions of the $c\to u,s$ transition form factors
have large differences between different methods. As indicated from
these two tables, results evaluated in
Refs.~\cite{Ivanov:2000aj,Kiselev:2002vz,Aliev:2006vs,Huang:2007kb}
are different with the other ones and ours to a large extent. In
Ref.~\cite{Kiselev:2002vz}, all of the results except for the $B_c$
to charmonium transitions are larger than the other results: the
authors have taken  into account the $\alpha_s/v$ corrections and
the form factors are enhanced by three times due to the Coulomb
renormalization of quark-meson vertex for the heavy quarkonium
$B_c$. Moreover, small decay constants for the $B$ meson are adopted
which also give large form factors: $f_B=140-170$ MeV,
$f_{B^*}/f_B=1.11$ and $f_{B_s}/f_B=1.16$. In
Ref.~\cite{Huang:2007kb}, the authors have chosen the chiral
correlation functions to derive the form factors in the light-cone
sum rules. Although only the twist-2  distribution amplitudes (DAs)
contribute and contributions from the twist-3 DAs vanish,
uncertainties of the continuum and the higher resonance interpolated
by both of the axial-vector current and vector current are expected
to be larger. In Ref.~\cite{Aliev:2006vs}, the authors also adopted
the three-point QCD sum rules but different correlation functions
are chosen. The form factors $A_2^{B_c B^*}$ and $A_2^{B_c B^*_s}$
in Ref.~\cite{Ivanov:2000aj} have different signs with the other
results. The large differences  in different models  can be used to
distinguish them in the future.


\begin{table}
\caption{$B_c\to D,D^*$ and $B_c\to D_s,D^*_s$ form factors at
$q^2=0$ evaluated in the literature. }\label{tab:formfactorsD}
\begin{tabular}{c|cccccccccc}
 \hline\hline
    & $F_1^{B_c D}=F_0^{B_c D}$ & $A_0^{B_c D^*}$ & $A_1^{B_c D^*}$ & $A_2^{B_c D^*}$ &$V^{B_c D^*}$ \\
 \hline
    DW\cite{Du:1988ws}\footnote{We quote the results with $\omega=0.6$ GeV.}
    & $0.154$ & $0.156$ & $0.145$ & $0.134$  & $0.224$ \\
 \hline
    CNP\cite{Colangelo:1992cx}
     & $0.13$  & $0.05$ &$0.11$  & $0.17$  & $0.25$ \\\hline
    NW\cite{Nobes:2000pm}
     & $0.1446$ & $0.094$ &$0.100$ &$0.105$ &$0.175$\\
 \hline
    IKS\cite{Ivanov:2000aj}
     & $0.69$ & $0.47$  &$0.56$  & $0.64$ & $0.98$ \\
 \hline
    Kiselev\cite{Kiselev:2002vz}\footnote{  The results out (in) the brackets are evaluated in sum rules (potential model).}
     & $0.32[0.29]$ & $0.35[0.37]$  &$0.43[0.43]$  & $0.51[0.50]$ & $1.66[1.74]$ \\
 \hline
    EFG\cite{Ebert:2003cn}
     & $0.14$ & $0.14$  &$0.17$  & $0.19$ & $0.18$ \\
 \hline
    HZ\cite{Huang:2007kb}
     & $0.35$ &  $0.05$ &$0.32$  & $0.57$ & $0.57$ \\
 \hline
    DSV\cite{DSV}
     & $0.075$ &  $0.081$ &$0.095$  & $0.11$ & $0.16$ \\
    \hline\hline
    & $F_1^{B_c D_s}=F_0^{B_c D_s}$ & $A_0^{B_c D_s^*}$ & $A_1^{B_c D^*_s}$
    & $A_2^{B_c D^*_s}$ &$V^{B_c D^*_s}$ \\
 \hline
    Kiselev\cite{Kiselev:2002vz}$^b$
     & $0.45[0.43]$ & $0.47[0.52]$  &$0.56[0.56]$  & $0.65[0.60]$ & $2.02[2.27]$ \\
    DSV\cite{DSV}
     & $0.15$ &  $0.16$ &$0.18$  & $0.20$ & $0.29$ \\%
    \hline\hline
\end{tabular}
\end{table}

\begin{table}
\caption{$B_c\to \eta_c, J/\psi,B,B^*,B_s,B_s^*$ form factors at
$q^2=0$ evaluated in the literature.
}\label{tab:formfactorcharmonium}
\begin{tabular}{c|cccccccccc}
 \hline\hline
    & $F_1^{B_c \eta_c}=F_0^{B_c \eta_c}$ & $A_0^{B_c J/\psi}$ & $A_1^{B_c J/\psi}$
    & $A_2^{B_c J/\psi}$   & $V^{B_c J/\psi}$ \\
 \hline
    DW\cite{Du:1988ws}\footnote{We quote the results with $\omega=0.6$ GeV.}
    & $0.420$ & $0.408$ & $0.416$ & $0.431$  & $0.591$ \\
 \hline
    CNP\cite{Colangelo:1992cx}
     & $0.20$  & $0.26$ &$0.27$  & $0.28$  & $0.38$ \\
 \hline
    KT\cite{Kiselev:1993ea}
     & $0.23$  & $0.21$ &$0.21$  & $0.23$  & $0.33$ \\
 \hline
    KLO\cite{Kiselev:1999sc}\footnote{
    We quote the values where the Coulomb corrections are taken into account. }
     & $0.66$  & $0.60$ &$0.63$  & $0.69$  & $1.03$ \\\hline
    NW\cite{Nobes:2000pm}
     & $0.5359$ & $0.532$ &$0.524$  &$0.509$ & $0.736$\\
 \hline
    IKS\cite{Ivanov:2000aj}
     & $0.76$ & $0.69$  &$0.68$  & $0.66$ & $0.96$ \\
 \hline
    Kiselev\cite{Kiselev:2002vz}\footnote{The results out (in) the brackets are evaluated in sum rules (potential model). }
     & $0.66[0.7]$ & $0.60[0.66]$  &$0.63[0.66]$  & $0.69[0.66]$ & $1.03[0.94]$ \\
 \hline
    EFG\cite{Ebert:2003cn}
     & $0.47$ & $0.40$  &$0.50$  & $0.73$ & $0.49$ \\
 \hline
    IKS2\cite{Ivanov:2005fd}
     & $0.61$ & $0.57$  &$0.56$  & $0.54$ & $0.83$ \\
 \hline
    HNV\cite{Hernandez:2006gt}
     & $0.49$ & $0.45$  &$0.49$  & $0.56$ & $0.61$ \\
 \hline
    HZ\cite{Huang:2007kb}
     & $0.87$ &  $0.27$ &$0.75$  & $1.69$ & $1.69$ \\
    \hline
    SDY\cite{Sun:2008ew}
     & $0.87$ &  $0.27$ &$0.75$  & $1.69$ & $1.69$ \\\hline
    DSV\cite{DSV}
     & $0.58$ &  $0.58$ &$0.63$  & $0.74$ & $0.91$ \\
    \hline\hline
    & $F_1^{B_c B}=F_0^{B_c B}$ & $A_0^{B_c B^*}$ & $A_1^{B_c B^*}$
    & $A_2^{B_c B^*}$ &$V^{B_c B^*}$ \\
 \hline
    DW\cite{Du:1988ws}$^{a}$
    & $0.662$ & $0.682$ & $0.729$ & $1.240$  & $5.690$ \\
 \hline
    CNP\cite{Colangelo:1992cx}
     & $0.3$   & $0.35$ & $0.34$ &$0.23$  & $1.97$ \\\hline
    NW\cite{Nobes:2000pm}
     & $0.4504$ & $0.269$ &$0.291$ &$0.538$&$1.94$ \\
 \hline
    IKS\cite{Ivanov:2000aj}\footnote{We add a minus sign to the form
    factors $F_1,A_0,A_1,A_2$ }
     & $0.58$ & $0.35$  &$0.27$  & $-0.60$ & $3.27$ \\
 \hline
    Kiselev\cite{Kiselev:2002vz}$^c$
     & $1.27[1.38]$ & $0.55[0.51]$  &$0.84[0.81]$  & $4.06[4.18]$ & $15.7[15.9]$ \\
 \hline
    EFG\cite{Ebert:2003cn}
     & $0.39$ & $0.20$  &$0.42$  & $2.89$ & $3.94$ \\
     \hline
    AS\cite{Aliev:2006vs}
     & ... & $0.28$  &$0.17$  & $-1.10$ & $0.09$ \\
     \hline
    HNV\cite{Hernandez:2006gt}\footnote{We add a minus sign for their predictions on the form factors.}
     & $0.39$ & $0.34$ & $0.38$ &$0.80$   & $1.69$ \\
 \hline
    HZ\cite{Huang:2007kb}
     & $0.90$ &  $0.27$ &$0.90$  & $7.9$ & $7.9$ \\\hline
    DSV\cite{DSV}
     & $0.41$ &  $0.42$ &$0.63$  & $2.74$ & $4.77$ \\
    \hline\hline
    & $F_1^{B_c B_s}=F_0^{B_c B_s}$ & $A_0^{B_c B^*_s}$ & $A_1^{B_c B^*_s}$
    & $A_2^{B_c B^*_s}$ &$V^{B_c B^*_s}$ \\
 \hline
    DW\cite{Du:1988ws}$^{a}$
    & $0.715$ & $0.734$ & $0.821$ & $1.909$  & $5.657$ \\
 \hline
    CNP\cite{Colangelo:1992cx}
     & $0.30$   & $0.39$ & $0.38$ &$0.35$  & $2.11$
     \\
 \hline
    CKM\cite{Choudhury:1998hs}\footnote{We quote the results which correspond to $m_b=4.9$ GeV and $\omega=0.4$ GeV. }
     & $0.403$   & $0.433$ & $0.487$ &$1.155$  & $3.367$
     \\
 \hline%
    NW\cite{Nobes:2000pm}
     & $0.5917$ & $0.445$ &$0.471$ &$0.787$ &$2.81$\\
 \hline
    IKS\cite{Ivanov:2000aj}$^d$
     & $0.61$ & $0.39$  &$0.33$  & $-0.40$ & $3.25$ \\\hline
    Kiselev\cite{Kiselev:2002vz}$^c$
     & $1.3[1.1]$ & $0.56[0.47]$  &$0.69[0.70]$  & $2.34[3.51]$ & $12.9[12.9]$ \\
 \hline
    EFG\cite{Ebert:2003cn}
     & $0.50$ & $0.35$  &$0.49$  & $2.19$ & $3.44$ \\
     \hline
    HNV\cite{Hernandez:2006gt}$^e$
     & $0.58$ & $0.52$& $0.55$   &$0.98$  & $2.29$ \\
 \hline
    HZ\cite{Huang:2007kb}
     & $1.02$ &  $0.36$ &$1.01$  & $9.04$ & $9.04$ \\\hline
    DSV\cite{DSV}
     & $0.55$ &  $0.57$ &$0.79$  & $3.24$ & $5.19$ \\
    \hline\hline
\end{tabular}
\end{table}

At the quark level, the $B_c\to P(V)l\bar\nu$ decays are described
as $b\to c(u) W^-\to c(u)l\bar\nu$ or $c\to d(s)W^+\to d(s) l^+\nu$.
Integrating out the highly offshell intermediate degrees of freedom
at tree level, the effective electroweak Hamiltonian for $b\to
ul\bar \nu_l$ transition, as an example, is
 \begin{eqnarray}
 {\cal H}_{eff}(b\to ul\bar \nu_l)=\frac{G_F}{\sqrt{2}}V_{ub}\bar
 u\gamma_{\mu}(1-\gamma_5)b \bar l\gamma^{\mu}(1-\gamma_5)\nu_l.
 \end{eqnarray}
Since  radiative corrections due to strong interactions only happen
between the $b$ quark and the $u$ quark, they characterize the
interactions at the low energy and  the Wilson coefficient which
contains the physics above the $m_b$ scale is not altered. With the
masses of leptons taken into account, the differential decay widths
of $B_c\to Pl\bar\nu$ and $B_c\to Vl\bar\nu$ ($l=e,\mu,\tau$) are
given by:
\begin{eqnarray}
 \frac{d\Gamma(B_c\to Pl\bar\nu)}{dq^2} &=&(\frac{q^2-m_l^2}{q^2})^2\frac{ {\sqrt{\lambda(m_{B_c}^2,m_P^2,q^2)}} G_F^2 |V_{\rm CKM}|^2} {384m_{B_c}^3\pi^3}
 \times \frac{1}{q^2} \nonumber\\
 &&\;\;\;\times \left\{ (m_l^2+2q^2) \lambda(m_{B_c}^2,m_P^2,q^2) F_1^2(q^2)  +3 m_l^2(m_{B_c}^2-m_P^2)^2F_0^2(q^2)
 \right\},\\
 \frac{d\Gamma_L(B_c\to Vl\bar\nu)}{dq^2}&=&(\frac{q^2-m_l^2}{q^2})^2\frac{ {\sqrt{\lambda(m_{B_c}^2,m_V^2,q^2)}} G_F^2 |V_{\rm CKM}|^2} {384m_{B_c}^3\pi^3}
 \times \frac{1}{q^2} \left\{ 3 m_l^2 \lambda(m_{B_c}^2,m_V^2,q^2) A_0^2(q^2)+\right.\nonumber\\
 &&\;\;\times  \left.(m_l^2+2q^2) \left|\frac{1}{2m_V}  \left[
 (m_{B_c}^2-m_V^2-q^2)(m_{B_c}+m_V)A_1(q^2)-\frac{\lambda(m_{B_c}^2,m_V^2,q^2)}{m_{B_c}+m_V}A_2(q^2)\right]\right|^2
 \right\},\label{eq:decaywidthlon}\\
 \frac{d\Gamma^\pm(B_c\to
 Vl\bar\nu)}{dq^2}&=&(\frac{q^2-m_l^2}{q^2})^2\frac{ {\sqrt{\lambda(m_{B_c}^2,m_V^2,q^2)}} G_F^2 |V_{\rm CKM}|^2} {384m_{B_c}^3\pi^3}
 \times   \nonumber\\
 &&\;\;\times \left\{ (m_l^2+2q^2) \lambda(m_{B_c}^2,m_V^2,q^2)\left|\frac{V(q^2)}{m_{B_c}+m_V}\mp
 \frac{(m_{B_c}+m_V)A_1(q^2)}{\sqrt{\lambda(m_{B_c}^2,m_V^2,q^2)}}\right|^2
 \right\},
\end{eqnarray}
where the subscript $+(-)$ denotes the right-handed (left-handed)
states of vector mesons. $\lambda(m_{B_c}^2,
m_{i}^2,q^2)=(m_{B_c}^2+m_{i}^2-q^2)^2-4m_{B_c}^2m_i^2$ with
$i=P,V$. The combined transverse and total differential decay widths
are given by:
\begin{eqnarray}
 \frac{d\Gamma_T}{dq^2}= \frac{d\Gamma^+}{dq^2}+
 \frac{d\Gamma^-}{dq^2},\;\;\;
 \frac{d\Gamma}{dq^2}= \frac{d\Gamma_L}{dq^2}+
 \frac{d\Gamma_T}{dq^2}.
\end{eqnarray}

As we have mentioned in the above, the form factors $A_2^{B_cB^*}$
and $A_2^{B_cB^*_s}$ only give small contributions to semileptonic
$B_c$ decays. In these two channels,  the two small variables
$m_{B_c}^2-m_{V}^2$ and $q^2$ satisfy the inequality: $q^2\leq
q^2_{\rm max}=(m_{B_c}-m_{V})^2\ll(m_{B_c}-m_{V})(m_{B_c}+m_{V})=
m_{B_c}^2-m_{V}^2$. One can expand the decay width in terms of small
variables. The variable $\lambda(m_{B_c}^2,m_V^2,q^2)$ can be
expanded as:
$\lambda(m_{B_c}^2,m_V^2,q^2)=(m_B^2-m_{V}^2)^2-4(m_B^2+m_V^2)q^2+q^4\sim
(m_{B_c}^2-m_{V}^2)^2$. From Eq.~(\ref{eq:decaywidthlon}), we can
see that the contribution from $A_2$ to the longitudinal
differential decay width contains a factor of
$\lambda(m_{B_c}^2,m_V^2,q^2)$ while the $A_1$ term is of the order
$\sqrt {\lambda(m_{B_c}^2,m_V^2,q^2)}$. Numerical results show that
the ratio
$\frac{\lambda(m_{B_c}^2,m_{B^*}^2,q^2)}{(m_{B_c}+m_{B^*})^2(m_{B_c}^2-m_{B^*}^2-q^2)}$
and
$\frac{\lambda(m_{B_c}^2,m_{B_s^*}^2,q^2)}{(m_{B_c}+m_{B^*_s})^2(m_{B_c}^2-m_{B^*_s}^2-q^2)}$
is smaller than 0.083 and $0.075$ in the full region for $q^2$,
respectively. It implies that the form factors $A_2^{B_cB^*}$ and
$A_2^{B_cB^*_s}$ can be safely neglected in the decay
width~\footnote{In  $B_c\to (B^*,B_s^*)V$ decays, the analysis is
similar: $q^2$ is replaced by the mass square of the vector meson
$m_V^2$.}.

\begin{table}
\caption{Branching ratios (in units of $\%$) and polarizations
$\frac{\Gamma_L}{\Gamma_T}$ of $B_c\to Ml\nu$ decays. The first kind
of uncertainties is from the $B_c$ decay constants and the decay
constant of the final state mesons, while the second one is from the
quark masses. The last kind of uncertainties is from the decay width
of $B_c$ and the CKM matrix element $V_{ub}$. The mass difference
between an electron and a muon does not provide sizable effects in
$B_c\to D^{(*)}l\bar\nu$ and $B_c\to \eta_c(J/\psi)l\bar\nu$ decays,
but it does in $B_c\to B^{(*)}l\nu$ and $B_c\to B^{{(*)}}_sl\nu$
decays}\label{tab:BR}
\begin{tabular}{c|cccccc}
 \hline\hline
                    & $B_c\to De\bar\nu_e$ & $B_c\to D\mu\bar\nu_\mu$ & $B_c\to D\tau\bar\nu_\tau$
                    & $B_c\to \eta_ce\bar\nu_e$ & $B_c\to \eta_c\mu\bar\nu_\mu$     \\\hline
    ${\cal BR} $    & $0.0030^{+0.0005+0.0005+0.0007}_{-0.0004-0.0004-0.0007}$
                    & $0.0030^{+0.0005+0.0005+0.0007}_{-0.0004-0.0004-0.0007}$
                    & $0.0021^{+0.0003+0.0003+0.0005}_{-0.0003-0.0003-0.0005}$
                    & $0.67^{+0.04+0.04+0.10}_{-0.07-0.04-0.10}$
                    & $0.67^{+0.04+0.04+0.10}_{-0.07-0.04-0.10}$ \\
 \hline\hline
                    & $B_c\to \eta_c\tau\bar\nu_\tau$ & $B_c\to Be\bar\nu_e$ & $B_c\to B\mu\bar\nu_\mu$
                    & $B_c\to B_s e\bar\nu_e$ & $B_c\to B_s\mu\bar\nu_\mu$     \\\hline
    ${\cal BR} $    & $0.190^{+0.005+0.014+0.029}_{-0.012-0.013-0.029}$
                    & $0.109^{+0.014+0.013+0.017}_{-0.016-0.012-0.017}$
                    & $0.104^{+0.013+0.013+0.016}_{-0.015-0.012-0.016}$
                    & $1.49^{+0.10+0.15+0.23}_{-0.13-0.14-0.23}$
                    & $1.41^{+0.09+0.14+0.21}_{-0.12-0.14-0.21}$ \\
 \hline\hline
                    & $B_c\to D^*e\bar\nu_e$ & $B_c\to D^*\mu\bar\nu_\mu$ & $B_c\to D^*\tau\bar\nu_\tau$
                    & $B_c\to J/\psi e\bar\nu_e$ & $B_c\to J/\psi\mu\bar\nu_\mu$     \\\hline
    ${\cal BR} $    & $0.0045^{+0.0005+0.0010+0.0011}_{-0.0004-0.0008-0.0010}$
                    & $0.0045^{+0.0005+0.0010+0.0011}_{-0.0004-0.0008-0.0010}$
                    & $0.0027^{+0.0003+0.0006+0.0007}_{-0.0002-0.0005-0.0006}$
                    & $1.49^{+0.01+0.15+0.23}_{-0.03-0.14-0.23}$
                    & $1.49^{+0.01+0.15+0.23}_{-0.03-0.14-0.23}$ \\\hline
    $\frac{\Gamma_L}{\Gamma_T}$
                    & $0.68^{+0.02+0.02+0.00}_{-0.02-0.02-0.00}$
                    & $0.68^{+0.02+0.02+0.00}_{-0.02-0.02-0.00}$
                    & $0.70^{+0.01+0.02+0.00}_{-0.01-0.02-0.00}$
                    & $1.04^{+0.00+0.02+0.00}_{-0.00-0.02-0.00}$
                    & $1.04^{+0.00+0.02+0.00}_{-0.00-0.02-0.00}$ \\
 \hline\hline
                    & $B_c\to J/\psi\tau\bar\nu_\tau$ & $B_c\to B^*e\bar\nu_e$ & $B_c\to B^*\mu\bar\nu_\mu$
                    & $B_c\to B_s^* e\bar\nu_e$ & $B_c\to B_s^*\mu\bar\nu_\mu$     \\\hline
    ${\cal BR} $    & $0.370^{+0.002+0.042+0.056}_{-0.005-0.038-0.056}$
                    & $0.141^{+0.002+0.029+0.021}_{-0.004-0.026-0.021}$
                    & $0.134^{+0.002+0.028+0.020}_{-0.004-0.025-0.020}$
                    & $1.96^{+0.00+0.34+0.30}_{-0.03-0.32-0.30}$
                    & $1.83^{+0.00+0.32+0.28}_{-0.03-0.30-0.28}$ \\\hline
    $\frac{\Gamma_L}{\Gamma_T}$
                    & $0.81^{+0.01+0.01+0.00}_{-0.01-0.01-0.00}$
                    & $1.07^{+0.01+0.02+0.00}_{-0.01-0.03-0.00}$
                    & $1.06^{+0.01+0.02+0.00}_{-0.01-0.02-0.00}$
                    & $1.14^{+0.01+0.02+0.00}_{-0.01-0.02-0.00}$
                    & $1.11^{+0.01+0.01+0.00}_{-0.01-0.02-0.00}$ \\
    \hline\hline
\end{tabular}
\end{table}

Integrating the differential decay widths over the variable $q^2$,
one obtains  partial decay widths and polarization fractions. The
lifetime of the $B_c$ meson and the relevant CKM matrix elements are
used as~\cite{Amsler:2008zz}:
\begin{eqnarray}
 \tau_{B_c}=(0.46\pm0.07) ps,\;\;\; |V_{cb}|=41.2\times
  10^{-3},\;\;\; |V_{ub}|=(3.93\pm0.36)\times 10^{-3},\;\;\; |V_{cd}|=0.230,\,\,\,
 |V_{cs}|=0.973,\,\,
\end{eqnarray}
where the small uncertainties in the other CKM matrix elements are
neglected.  Our predictions of branching ratios and polarization
quantities $\frac{\Gamma_L}{\Gamma_T}$ in semileptonic $B_c$ decays
are given in table~\ref{tab:BR}. The three kinds of  uncertainties
are from: the decay constants of the $B_c$ meson and the meson in
the final state;  the constituent quark masses; the lifetime of
$B_c$ together with the CKM matrix elements. The first kind of
uncertainties in the $B_c\to (D,D_s,J/\psi) l\bar\nu$ decays is very
small, as the uncertainties in decay constants of $D$ and $J/\psi$
are small. The different mass between the electron and muon does not
have sizable effects on $b\to u,c$ semileptonic decays, but the
branching ratios of $c\to u,s$ transitions are altered by roughly
$5\%$. Branching ratios of $B_c\to Pl\bar\nu$ decays are smaller
than the corresponding $B_c\to Vl\bar\nu$ ones, partly because there
are three kinds of polarizations for vector mesons.Among the four
kinds of transitions at the quark level, there is an inequation in
chain:
\begin{eqnarray}
 {\cal BR}(B_c\to D^*l\nu)<{\cal BR}(B_c\to B^*l\nu)<{\cal BR}(B_c\to
 J/\psi l\nu)<{\cal BR}(B_c\to B_s^*l\nu),
\end{eqnarray}
where we have taken decays involving a vector meson as an example.
To understand this inequation, three points are essential.  The CKM
matrix elements for these four kinds of decays are given as:
\begin{eqnarray}
  |V_{ub}|\ll V_{cb}\ll |V_{cd}|\ll V_{cs}.
\end{eqnarray}
The form factors at zero-recoiling point roughly respect:
\begin{eqnarray}
 F(B_c\to D^*)<F(B_c\to J/\psi)\sim F(B_c\to B^*)\sim F(B_c\to B^*_s).
\end{eqnarray}
The phase spaces in $B_c\to D^*$ and $B_c\to J/\psi$ transitions are
much larger than those in $B_c\to B^*,B_s^*$ transitions, which can
compensate for the small CKM matrix element in $B_c\to J/\psi l\bar
\nu$ decay. These predictions will be tested at the ongoing and
forthcoming hadron colliders.


\section{Conclusion}
\label{conc}

Due to the rich data,  measurements on the CKM matrix elements are
becoming more and more accurate. $B_c$ meson decays provide another
promising place to continue the errand in $B$ meson decays. They
also offer a new window to explore the structure of weak
interactions. Although the $B_c$ meson can not be produced on the
two B factories, it has a promising prospect on the ongoing and
forthcoming hadron colliders. Because of these interesting features,
we have studied the $B_c$ transition form factors in the covariant
light-front quark model, which are relevant for the semileptonic
$B_c$ decays.

Comparing our predictions with results for the form factors in the
literature, we find large discrepancies which may be useful to
distinguish various theoretical methods. Our results for the form
factors $A_2$ in $B_c\to B^*$ and $B_c\to B_s^*$ transitions
strongly depend on the decay constants of the $B^*$ and $B_s^*$
mesons, which gives large theoretical uncertainties to the form
factors. For $B_c\to BP$ decays, the relevant form factor $A_0$ is
almost independent of $A_2$: $A_0\simeq A_1$. For semileptonic $B_c$
decays (also $B_c\to B^*V$ decays), contributions from $A_2$ are at
least suppressed by a factor of $0.08$ compared with those from
$A_1$. Thus the large uncertainties from $A_2$ will not affect the
physical observables.

$B_c\to D,D^*,D_s,D_s^*$ form factors at maximally recoiling point
are smaller than $B_c\to \eta_c, J/\psi, B,B^*,B_s,B_s^*$, while the
$B_c\to D,D_s,\eta_c$ form factors at zero recoiling point are close
to each other.  The SU(3) symmetry breaking effects in $B_c\to
D,D_s$ and $B_c\to D^*,D_s^*$ are quite large; but in $B_c\to B,B_s$
and $B_c\to B^*,B_s^*$ transitions, the SU(3) breaking effects are
not large. Semileptonic $B_c\to (\eta_c,J/\psi)l\nu$ and $B_c\to
(B_s,B_s^*)l\nu$ decays have much larger branching fractions than
the other two kinds of semileptonic $B_c$ decays. In the three kinds
of $B_c\to Vl\nu$ decays, contributions from the longitudinal
polarized vector is comparable with those from the transversely
polarized vector.  These predictions will be tested at the ongoing
and forthcoming hadron colliders.

\section*{Acknowledgement}

This work is partly supported by the National Natural Science
Foundation of China under Grant Numbers 10735080, 10625525, and
10805037. We would like to acknowledge J.F. Sun and F. Zuo for
useful discussions.

\appendix
\section{Relations of Different definitions of form factors}\label{sec:definitionsofformfactors}

In the literature, various conventions for the $B_c\to V$ form
factors have been adopted. In this appendix, we will collect their
conventions and compare them with the BSW form factors.  In
Refs.~\cite{Colangelo:1992cx,Kiselev:1993ea,Kiselev:1999sc,Kiselev:2002vz}
, the authors defined the $B_c\to V$ form factors as:
\begin{eqnarray}
 \langle V(P'',\epsilon'')|V_\mu |\bar B_c(P')\rangle &=&-\epsilon_{\mu\nu\alpha \beta}\vp^{\pp*\nu}P^\alpha
    q^\beta  F_V(q^2),\\
 \langle V(P'',\epsilon'')|A_\mu |\bar B_c(P')\rangle &=&iF_0(q^2)\epsilon^{\prime\prime*}_\mu
 +iF_+(q^2)(\epsilon^{\prime\prime*}\cdot P)P_\mu
 +iF_-(q^2)(\epsilon^{\prime\prime*}\cdot P)q_\mu,
\end{eqnarray}
These form factors are related to the BSW form factors by:
\begin{eqnarray}
 V^{PV}&=&(m_{B_c}+m_V)F_V,\;\;\; A_1^{PV}=\frac{F_0}{m_{B_c}+m_V},\;\;\;
 A_2^{PV}=-(m_{B_c}+m_V)F_+,\\
  A_0&=&\frac{m_{B_c}+m_{V}}{ 2m_{V}}
 A_1^{PV}(q^2)-\frac{m_{B_c}-m_{V}}{2m_{V}}\,A_2^{PV}(q^2)+\frac{q^2}{2m_V}F_-.
\end{eqnarray}
The definition of form factors $g,f,a_+,a_-$ in
Ref.~\cite{Nobes:2000pm} is similar with ours in
Eqs.(\ref{eq:formfactor1}-\ref{eq:formfactor3}) except for a phase
$i$. In Ref.~\cite{Ivanov:2005fd,Hernandez:2006gt}, the following
definition for the form factors is adopted:
\begin{eqnarray}
 \langle V(P'',\epsilon'')|V_\mu-A_\mu|\bar B_c(P')\rangle
 &=&\frac{i}{m_{B_c}+m_V}\epsilon^{\prime\prime*}_\nu\left(-g^{\mu\nu}P\cdot q A_0
 +P^\mu P^\nu A_++q^\mu P^\nu A_-+i\epsilon^{\mu\nu\rho\sigma}P_\rho q_\sigma
 V\right),
\end{eqnarray}
where $A_+$ corresponds to the BSW form factor $A_2^{PV}$ and their
form factor $A_0^{IKS2}$ is related to the BSW form factor
$A_1^{PV}$:
\begin{eqnarray}
 A_1^{PV}=\frac{A_0^{IKS2}(m_{B_c}-m_V)}{m_{B_c}+m_V}.
\end{eqnarray}
In Ref.~\cite{Huang:2007kb}, the $B_c\to V$ form factors are defined
as:
\begin{eqnarray}
 \langle V(P'',\epsilon'')|V_\mu-A_\mu|\bar B_c(P')\rangle
 &=& -i\epsilon^{\prime\prime*}_\nu (m_{B_c}+m_V)A_1+iP_\mu(\epsilon^{\prime\prime*}\cdot q)\frac{A_+}{m_{B_c}+m_V}
 +iq_\mu(\epsilon^{\prime\prime*}\cdot
 q)\frac{A_-}{m_{B_c}+m_V}\nonumber\\
 &&
 +\epsilon_{\mu\nu\rho\sigma}\epsilon^{\prime\prime*}_\nu q_\rho P_\sigma \frac{V}{m_{B_c}+m_V},
\end{eqnarray}
The  form factors $A_1^{PV}$ and $V^{PV}$ are the same with the
relevant BSW form factors; their form factor $A_+$ corresponds to
the BSW form factor $A_2^{PV}$.

\section{Some specific rules under the $p^-$ integration}\label{sec:rules}

When performing the $p^-$ integration, one needs to include the
zero-mode contribution. This amounts to performing the integration
in a proper way in this approach. To be more specific, for $\hat
p_1^\prime $ under integration we use the following
rules~\cite{Jaus:1999zv,Cheng:2003sm}
 \begin{eqnarray}
\hat p^\prime_{1\mu}
       &\doteq& P_\mu A_1^{(1)}+q_\mu A_2^{(1)},
\hat N_2
       \to Z_2,
  \nonumber\\
\hat p^\prime_{1\mu} \hat p^\prime_{1\nu}
       &\doteq& g_{\mu\nu} A_1^{(2)}+P_\mu P_\nu A_2^{(2)}+(P_\mu
                q_\nu+ q_\mu P_\nu) A^{(2)}_3+q_\mu q_\nu A^{(2)}_4,%
 \label{eq:p1B}
 \end{eqnarray}
where the symbol $\doteq$ reminds us that the above equations are
true only after integration.  $A^{(i)}_j$ are functions of
$x_{1,2}$, $p^{\prime2}_\bot$, $p^\prime_\bot\cdot q_\bot$ and
$q^2$, and their explicit expressions have been studied in
Ref.~\cite{Jaus:1999zv,Cheng:2003sm}:
 \begin{eqnarray}
 Z_2&=&\hat N_1^\prime+m_1^{\prime2}-m_2^2+(1-2x_1)M^{\prime2}
 +(q^2+q\cdot P)\frac{p^\prime_\bot\cdot q_\bot}{q^2},  \nonumber\\
  A^{(1)}_1&=&\frac{x_1}{2},
 \quad%
 A^{(1)}_2=A^{(1)}_1-\frac{p^\prime_\bot\cdot q_\bot}{q^2},
 A^{(2)}_1=-p^{\prime2}_\bot-\frac{(p^\prime_\bot\cdot q_\bot)^2}{q^2},
 \quad%
 \nonumber\\
 A^{(2)}_3&=&A^{(1)}_1 A^{(1)}_2,
 A^{(2)}_4=\big(A^{(1)}_2\big)^2-\frac{1}{q^2}A^{(2)}_1.
 \quad%
 \label{eq:rule}
 \end{eqnarray}
We do not show the spurious contributions in Eq.~(\ref{eq:rule})
since they are numerically vanishing.


\end{document}